\documentclass[12pt,a4paper]{article}

\usepackage{amsmath}
\usepackage{amssymb}
\usepackage{graphicx}
\usepackage{dsfont}
\usepackage{overpic}
\usepackage{cite}
\usepackage{slashed}

\let\oldappendix=\appendix
\let\oldsection=\section
\renewcommand{\appendix}{\oldappendix%
\def\theequation{\Alph{section}.\arabic{equation}}%
\renewcommand{\section}{\setcounter{equation}{0}\oldsection}}

\newcommand{\beq}{\begin{equation}}
\newcommand{\eeq}{\end{equation}}
\newcommand{\beqa}{\begin{eqnarray}}
\newcommand{\eeqa}{\end{eqnarray}}

\newcommand{\tr}{\mbox{tr}}

\newcommand{\newop}[2]{\def#1{\mathop{\mathrm{#2}}\nolimits}}
\newop{\artanh}{artanh}
\newop{\det}{det}
\newop{\tr}{tr}
\newop{\diag}{diag}
\newop{\Re}{Re}
\newop{\Im}{Im}

\begin{document}

%\hfill 

%\hfill 

\bigskip
\bigskip

\begin{center}

{{\Large\bf  The triton and three-nucleon force\\[0.3cm] in nuclear lattice simulations }}

\end{center}

\vspace{.4in}

\begin{center}
{\large B.~Borasoy\footnote{email: borasoy@itkp.uni-bonn.de}$^a$,
        H.~Krebs\footnote{email: hkrebs@itkp.uni-bonn.de}$^a$,
        D.~Lee\footnote{email: dean\_lee@ncsu.edu}$^b$,
        U.-G.~Mei{\ss}ner\footnote{email: meissner@itkp.uni-bonn.de}$^{a,c}$}

\hfill 

\hfill

\vspace{.4in}

$^a$Helmholtz-Institut f\"ur Strahlen- und Kernphysik (Theorie) \\
Universit\"at Bonn, Nu{\ss}allee 14-16, D-53115 Bonn, Germany \\[0.3cm]
$^b$North Carolina State University, Raleigh, NC 27603, USA \\[0.3cm]
$^c$Institut f\"ur Kernphysik (Theorie), Forschungszentrum J\"ulich \\
D-52425 J\"ulich, Germany \\

\vspace{.2in}

\end{center}

\vspace{.7in}

\thispagestyle{empty}

\begin{abstract}
We study the triton and three-nucleon force at lowest chiral order in pionless
effective field theory both in the Hamiltonian and Euclidean nuclear lattice formalism.
In the case of the Euclidean lattice formalism, we derive the exact few-body
worldline amplitudes corresponding to the standard many-body lattice action.
This will be useful for setting low-energy coefficients in future nuclear lattice simulations.  
We work in the
Wigner SU(4)-symmetric limit where the $S$-wave scattering lengths $^{1}S_{0}$ and $^{3}S_{1}$
are equal. By comparing with continuum 
results, we demonstrate for the first time that the nuclear lattice 
formalism can be used to study few-body nucleon systems. 

\end{abstract}

\vfill

%%%%%%%%%%%%%%%%%%%%%%%%%%%%%%%%%%%%%%%%%%%%%%%%%%%%%%%%%%%%%%%%%%%%%%%%%%%%%%%%%
\section{Introduction}

The ultimate goal of nuclear physics is the derivation of physical observables from the
underlying theory of the strong interactions, quantum chromodynamics (QCD).
At low energies quarks and gluons are confined in nucleons and pions and
one is forced to work in the non-perturbative regime of QCD.
A model-independent approach to non-perturbative QCD is provided by lattice
QCD simulations. However, modern lattice QCD simulations work with lattices not
much larger than the size of a single nucleon. In the foreseeable future lattice
QCD calculations will therefore not be suited to obtain direct results
in few-body or many-body nuclear physics.

Alternatively, one can work with the effective degrees of freedom at such energies,
i.e., the nucleons and pions. Their interactions are summarized in an effective chiral
Lagrangian which in turn may be utilized, e.g., as the interaction kernel
in the Lippmann-Schwinger or Faddeev-Yakubovsky equations for few-nucleon systems. 
The advantage of the effective chiral Lagrangian approach with respect to traditional
potential models is the
clear theoretical connection to QCD and the possibility to improve the
calculations systematically by going to higher chiral orders.
Over the last few years, effective field theory methods have been applied
to few-nucleon systems, see e.g. \cite{Bedaque:2002mn, Epelbaum:2005pn} and references therein. 
In particular, in the three-nucleon system
this framework provided an explanation for universal features such as the Phillips line \cite{Phillips},
the Efimov effect \cite{Efimov:1971a, Efimov:1993a}, and the Thomas effect \cite{Thomas}, which
have no obvious explanation in conventional potential models \cite{Braaten:2004rn}.

Very recently, chiral effective field theory methods on the lattice have been applied to 
nuclear and neutron matter \cite{Lee:2004si, Lee:2004qd, Lee:2005is, Lee:2005it}. The chiral effective
Lagrangian is discretized on the lattice and the path integral is evaluated
non-perturbatively using Monte Carlo sampling.
In this work we employ pionless effective field theory at lowest chiral order on the lattice
to study the triton and three-nucleon force.
We work in the Wigner SU(4)%
-symmetric limit where the $^{1}S_{0}$ and the $^{3}S_{1}$ scattering lengths
are the same, and the isospin-spin SU(2)$\times$SU(2) symmetry is promoted to
an SU(4) symmetry. 

This is the first study of the three-nucleon system in the lattice formalism
and the motivations for this study are twofold.  First, it provides a starting
point for more realistic studies of the triton system as well as other
few-nucleon systems using effective field theory and efficient Monte
Carlo lattice methods. Second, it provides the first measurement of the
three-nucleon force on the lattice in the Wigner symmetry limit.  This
provides an important step towards future many-body nuclear simulations with
arbitrary numbers of neutrons and protons including three-body effects.

This work is organized as follows. We start in the next section by reviewing
continuum regularization results for the triton binding energy. This will
serve us as a reference for the lattice calculations in the subsequent sections.
In Sec.~\ref{sec:lanczos} we consider the Hamiltonian lattice using the Lanczos
method, while in Sec.~\ref{sec:euclat} the Euclidean lattice formulation
is utilized  and the path integral is evaluated with Monte Carlo sampling.
Our results indicate that many-body simulations of dilute nuclear matter 
in the Wigner symmetry limit should be
possible without a sign problem, even for unequal numbers of protons and neutrons.
This will be discussed in Sec.~\ref{sec:hubstr}, while our conclusions are
presented in Sec.~\ref{sec:concl}.
A few formulae are deferred to the Appendix.

%%%%%%%%%%%%%%%%%%%%%%%%%%%%%%%%%%%%%%%%%%%%%%%%%%%%%%%%%%%%%%%%%%%%%%%%%%%%%%%

\section{Continuum regularization} \label{sec:contreg}

The nontrivial dependence of the three-body force on the cutoff is a
non-perturbative effect.  There is no finite set of diagrams which produces
the ultraviolet divergence.  
According to the Thomas effect \cite{Thomas} a two-body interaction with range $R$ and fixed
binding energy $B_2$ produces in the zero range limit a deeply bound three-body
state with a binding energy that scales as $1/R^2$, see \cite{Braaten:2004rn} for a review. 
This singular behavior of
the three-body system yields a three-body force that
could be quite a different function of the cutoff scale
for different regularization schemes.  In this section we therefore consider first the usual
continuum regularization which will then be compared to lattice regularization employed 
in the following sections.

\begin{figure}
[ptb]
\begin{center}
\includegraphics[
height=3.in,
width=3.8in%,
%angle=-90
]%
{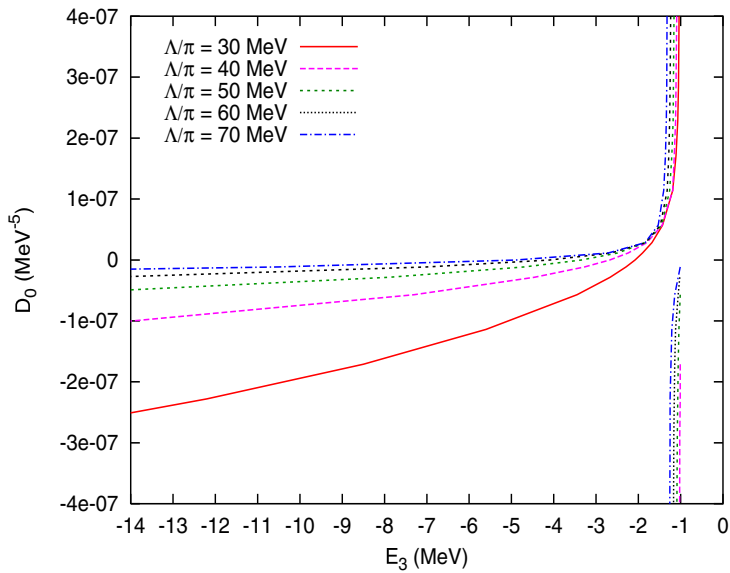}%
\caption{The three-body interaction $D_{0}$ as a function of the triton energy
$E_{3}$ for various cutoff momenta using continuum regularization. \ The
two-nucleon binding energy is set at $1$ MeV.}%
\label{equation7_d0}%
\end{center}
\end{figure}

In the SU(4)-symmetric limit, where the isospin and spin degrees of freedom
can be interchanged, the lowest order non-relativistic effective
Lagrange density including three-body interactions reads%
\begin{equation} \label{eq:Lagr}
L=\psi^{\dagger}\left(  i\partial_{0}+\frac{\vec{\nabla}^{2}}{2m}\right)
\psi-\frac{C_{0}}{2}\left(  \psi^{\dagger}\psi\right)  ^{2}-\frac{D_{0}}%
{6}\left(  \psi^{\dagger}\psi\right)  ^{3},
\end{equation}
where $\psi$ is a four-component column vector of the four nucleon states with mass $m$,%
\begin{equation}
\psi=\left[
\begin{array}
[c]{c}%
p_{\uparrow}\\
p_{\downarrow}\\
n_{\uparrow}\\
n_{\downarrow}%
\end{array}
\right]  .
\end{equation}
After summing bubble diagrams in the dinucleon or dimer propagator and
renormalizing the two-body interaction, we have for the renormalized value of $C_0$ the relation%
\begin{equation}
C_{0}=\frac{4\pi a_{2}}{m},
\end{equation}
where $a_{2}$ is the two-body scattering length.  Following
\cite{Bedaque:1998km, Bedaque:1999ve}, the homogeneous $S$-wave bound state
equation for dimer-nucleon scattering is%
\begin{equation}
t(k,p)=\frac{2}{\pi}\int_{0}^{\Lambda}dq\frac{t(k,q)q^{2}}{-\frac{1}{a_{2}%
}+\sqrt{\frac{3q_{{}}^{2}}{4}-mE_3}}\left[  \frac{1}{pq}\ln\left(  \frac
{p^{2}+pq+q^2-mE_3}{p^{2}-pq+q^{2}-mE_3}\right)  -\frac{D_{0}}{3mC_{0}^{2}}\right]
, \label{integral eq}%
\end{equation}
where $\Lambda$ is the cutoff momentum, $E_3$ is the total energy, and $k$ is determined by
\begin{equation}
k=\sqrt{\frac{4m}{3}(E_3+B_{2})} \ ,
\end{equation}
with $B_{2}$ the two-body binding
energy%
\begin{equation}
B_{2}=\frac{1}{ma_{2}^{2}} \ \text{.}%
\end{equation}
In the $SU(4)$-symmetric limit, the $^1S_0$ and $^3S_1$ two-body sectors are degenerate.
Since the $^1S_0$ channel has a nearly zero energy resonance,
we take the $SU(4)$-symmetric two-nucleon binding energy to be about one-half the
deuteron binding,
\begin{equation}
B_{2}=1\text{ MeV}.
\end{equation}
The coupling $D_0$ is shown in Fig. \ref{equation7_d0} as a function of $E_3$.  In the notation of
\cite{Bedaque:1998km} we have%
\begin{equation}
H(\Lambda)=\frac{h\Lambda^{2}}{2mg^{2}}=-\frac{D_{0}\Lambda^{2}}{6mC_{0}^{2}%
}=-\frac{D_{0}m\Lambda^{2}}{96\pi^{2}a_2^2},
\end{equation}
with $H(\Lambda)\sim1$ away from critical values of $\Lambda$,  
and so $D_{0}$ scales roughly as $\Lambda^{-2}$ for $E_3$ much less than $-1$~MeV, as
can be seen in Fig. \ref{equation7_d0}.%
%TCIMACRO{\FRAME{ftbpFU}{2.981in}{4.2436in}{0pt}{\Qcb{The three-body
%interaction $D_{0}$ as a function of the triton energy $E_{3}$ for various
%cutoff momenta using continuum regularization.  The two-nucleon binding
%energy is set at $1$ MeV.}}{\Qlb{equation7_d0}}{equation7_d0.eps}%
%{\special{ language "Scientific Word";  type "GRAPHIC";
%maintain-aspect-ratio TRUE;  display "USEDEF";  valid_file "F";
%width 2.981in;  height 4.2436in;  depth 0pt;  original-width 4.8948in;
%original-height 6.9877in;  cropleft "0";  croptop "1";  cropright "1";
%cropbottom "0";
%filename 'equation7_D0.eps';file-properties "XNPEU";}}}%
%BeginExpansion
%EndExpansion
We observe a pole in $D_{0}$ as a function of $E_{3}$.  For the cutoff
values considered here the pole occurs rather close to the $-1$ MeV continuum
threshold for a dimer plus nucleon.  The location of this pole,
$E_{3}^{\text{pole}}$, slowly increases in magnitude with the cutoff $\Lambda
$. For $E_{3}$ greater than $E_{3}^{\text{pole}}$, there is no way to
accommodate the triton as the ground state of the three-body system.  As we
cross $E_{3}^{\text{pole}}$ we can instead switch labels and identify the
triton with the first excited $S$-wave state.  If we do this then the value of
$D_{0}$ must switch from infinitely repulsive to infinitely attractive, and
the new ground state now lies outside the range of validity of effective field
theory.  
%This deeply bound state becomes a so-called Efimov state
%\cite{Efimov:1971a, Efimov:1993a} in the limit $\Lambda a_{2} \rightarrow\infty$.  
For very large $\Lambda a_{2}$ we expect to see many
such poles, but in our case $\Lambda a_{2}$ is less than $10$ and there is
only one pole.

%%%%%%%%%%%%%%%%%%%%%%%%%%%%%%%%%%%%%%%%%%%%%%%%%%%%%%%%%%%%%%%%%%%%%%%%%%%%%%%
\section{Hamiltonian lattice with Lanczos method} \label{sec:lanczos}

In this section we consider the triton and three-body force in the Hamiltonian
lattice formalism.  After constructing the lattice Hamiltonian for the
three-nucleon system, we use the Lanczos method \cite{Lanczos} to find the lowest
eigenvalues. Since we compute several low eigenvalues, the Lanczos method is well-suited to observing
the singular behavior of the three-body coupling $D_0$ in the triton energy. 
However, the method requires storing in memory the space of three-nucleon states at rest.
Thus we cannot probe larger volumes nor any useful sytems with more than three particles.

Let $a_{i}(\vec{n}_{s})$ be an annihilation operator for a
nucleon with spin-isospin index $i$ at the spatial lattice site $\vec{n}_{s}$. \ The
lattice Hamiltonian with two-body and three-body interactions corresponding
to the Lagrangian in Eq.~(\ref{eq:Lagr}) is%
\begin{align}  \label{eq:hamil}
H &  =\frac{3}{m}\sum_{\vec{n}_{s},i}a_{i}^{\dagger}(\vec{n}_{s})a_{i}%
(\vec{n}_{s})
 -\frac{1}{2m}\sum_{\vec{n}_{s},i}\sum_{l_{s}=1,2,3}\left[  a_{i}^{\dagger
}(\vec{n}_{s})a_{i}(\vec{n}_{s}+\hat{l}_{s})+a_{i}^{\dagger}(\vec{n}_{s}%
)a_{i}(\vec{n}_{s}-\hat{l}_{s})\right]  \nonumber\\
&  +C_{0}\sum_{\vec{n}_{s},i>j}a_{i}^{\dagger}(\vec{n}_{s})a_{i}(\vec{n}%
_{s})a_{j}^{\dagger}(\vec{n}_{s})a_{j}(\vec{n}_{s})%\nonumber\\
%&  
+D_{0}\sum_{\vec{n}_{s},i>j>k}a_{i}^{\dagger}(\vec{n}_{s})a_{i}(\vec{n}%
_{s})a_{j}^{\dagger}(\vec{n}_{s})a_{j}(\vec{n}_{s})a_{k}^{\dagger}(\vec{n}%
_{s})a_{k}(\vec{n}_{s})
\end{align}
with the unit lattice vectors $\hat{l}_{s}$ in the spatial directions.
All parameters are given in lattice units, i.e., physical parameters are multiplied by
the appropriate power of the lattice spacing $a$.  As derived in
\cite{Lee:2005it}, $C_{0}$ is determined by the nucleon-nucleon scattering
length $a_{2}$,%
\begin{equation}
C_{0}=\frac{1}{m}\frac{1}{\frac{1}{4\pi a_{2}}-\frac{1}{L^{3}}\sum\limits_{\vec
{k}\neq0\text{ }\operatorname{integer}}\frac{1}{\Omega_{_{\vec{k}}}}},
\end{equation}
where%
\begin{equation}
\Omega_{_{\vec{k}}}=6-2\cos\frac{2\pi k_{1}}{L}-2\cos\frac{2\pi k_{2}}%
{L}-2\cos\frac{2\pi k_{3}}{L}.
\end{equation}

One disadvantage of using lattice discretization is that rotation invariance
is no longer exact.  We cannot project out $S$-wave bound states as in standard
continuum regularization.  Nevertheless we can identify the subspace of
three-nucleon states with zero total momentum and construct the corresponding
Hamiltonian submatrix.  We then use the Lanczos method to identify the lowest
energy eigenvalues and states invariant under the full cubic symmetry group of the lattice.

\begin{figure}
[hbt]
\begin{center}
\includegraphics[
height=3.in,
width=3.8in%,
%angle=-90
]%
{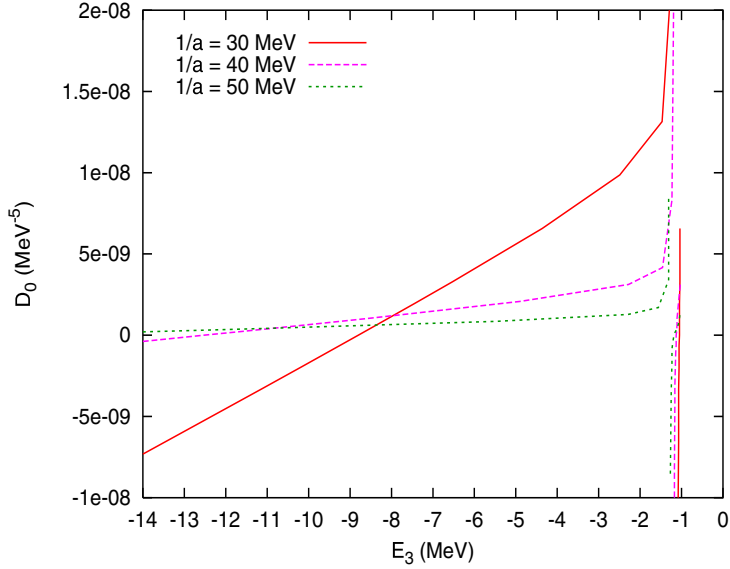}%
\caption{The three-body interaction $D_{0}$ as a function of the triton energy
$E_{3}$ for various lattice spacings using the Hamiltonian lattice formalism.
\ The two-nucleon binding energy is set at $1$~MeV.}%
\label{hamiltonian}%
\end{center}
\end{figure}

For lattice spacings $a^{-1}=$ $30,40,50$ MeV we consider lattices with length
$L=4,5,6,7$. The dimension of the space is given by $L^6$ which can be seen as follows:
in the rest frame, we need only to specify the relative displacements of
the three distinguishable particles.  This is given by two vectors and
hence $(L^3)^2 = L^6$ possibilities.
We then use these finite volume results to extrapolate to the infinite
volume limit. 
\ The results are shown in Fig. \ref{hamiltonian}.%
%TCIMACRO{\FRAME{ftbpFU}{2.981in}{4.2436in}{0pt}{\Qcb{The three-body
%interaction $D_{0}$ as a function of the triton energy $E_{3}$ for various
%lattice spacings using the Hamiltonian lattice formalism. \ The two-nucleon
%binding energy is set at $1$ MeV.}}{\Qlb{hamiltonian}}{hamiltonian.eps}%
%{\special{ language "Scientific Word";  type "GRAPHIC";
%maintain-aspect-ratio TRUE;  display "USEDEF";  valid_file "F";
%width 2.981in;  height 4.2436in;  depth 0pt;  original-width 4.8948in;
%original-height 6.9877in;  cropleft "0";  croptop "1";  cropright "1";
%cropbottom "0";
%filename 'Hamiltonian.eps';file-properties "XNPEU";}}}%
%BeginExpansion
%EndExpansion
As with the continuum regularization results we see that $D_{0}$ scales
roughly as $\Lambda^{-2}$.  In contrast with continuum regularization
results, however, it seems that $D_{0}$ is also shifted towards more positive
values for smaller values of $\Lambda$. There is again a pole singularity
not too far below the threshold value of $-1$ MeV and the location of this pole
again slowly increases in magnitude with the inverse lattice spacing.

%%%%%%%%%%%%%%%%%%%%%%%%%%%%%%%%%%%%%%%%%%%%%%%%%%%%%%%%%%%%%%%%%%%%%%%%%%%%%

\section{Euclidean lattice with Monte Carlo} \label{sec:euclat}

In this section we consider the triton and three-body force in the Euclidean
lattice formalism.  We employ the lattice action described in \cite{Lee:2004qd}
for pure neutron matter and extend it to include protons in the Wigner
SU(4) limit. Since the formalism was originally designed for many-body
simulations at fixed chemical potential and arbitrary number of particles,
we remove the dependence on the chemical potential.
We now go through the steps somewhat carefully to find the corresponding
Euclidean lattice formalism for fixed number of particles.  Our rationale for
this extra care is to preserve the same lattice discretization errors due to
the spatial and temporal lattice spacings as in \cite{Lee:2004qd}, so that
we can use the few-body results to fix coefficients in the many-body simulation.

\subsection{Free particle}

We start by studying the lattice action for a free single fermion.
On the lattice the free Hamiltonian for a single fermion species is given by, see Eq.~(\ref{eq:hamil}),
\beq
H^{free}   =\frac{3}{m}\sum_{\vec{n}_{s}}a^{\dagger}(\vec{n}_{s})a%
(\vec{n}_{s})
 -\frac{1}{2m}\sum_{\vec{n}_{s}}\sum_{l_{s}=1,2,3}\left[  a^{\dagger
}(\vec{n}_{s})a(\vec{n}_{s}+\hat{l}_{s})+a^{\dagger}(\vec{n}_{s}%
)a(\vec{n}_{s}-\hat{l}_{s})\right] \ .
\eeq
One can approximate the partition function as a Euclidean path integral
\beq  \label{eq:freepart}
Z^{free} = \tr \left[ \exp \left( - \beta H^{free} \right) \right] \simeq \int Dc \, 
                      Dc^* \exp \left( - S^{free} \right) \ ,
\eeq
where we have dropped an irrelevant constant on the left side of Eq.~(\ref{eq:freepart}).
The lattice action for the free non-relativistic fermion is given by%

\begin{align}
S^{free} &  =\sum_{\vec{n}}\left[  c^{\ast}(\vec{n})c(\vec{n}+\hat
{0})-(1-6h)c^{\ast}(\vec{n})c(\vec{n})\right]
\nonumber\\
&  -h\sum_{\vec{n},l_{s}}\left[  c^{\ast}%
(\vec{n})c(\vec{n}+\hat{l}_{s})+c^{\ast}(\vec{n})c(\vec{n}-\hat{l}%
_{s})\right]  
\end{align}
with $h = \alpha_t/(2 m)$ and $\alpha_t= a_t/a$ and Grassmann variables $c, c^*$,
while the $\hat{0}$ signifies one lattice unit in the forward time direction.
Note that the relation between the partition function and the Euclidean path integral in
Eq.~(\ref{eq:freepart}) is exact for $h=0$, so that discretizetion effects
in temporal direction are minimized.
By translating the $c$ field back by one unit in the time direction
and using the same notation for simplicity, we can write this in the form%
\begin{align}
S^{free} &  =\sum_{\vec{n}}\left[  c^{\ast}(\vec{n})c(\vec{n})-
(1-6h)c^{\ast}(\vec{n})c(\vec{n}-\hat{0})\right]  \nonumber\\
&  -h\sum_{\vec{n},l_{s}}\left[  c^{\ast}%
(\vec{n})c(\vec{n}+\hat{l}_{s}-\hat{0})+c^{\ast}(\vec{n})c(\vec{n}-\hat
{l}_{s}-\hat{0})\right] \ . 
\end{align}

\begin{figure}[hbt]
\begin{center}
\includegraphics[height=2.3in,width=2.5in,angle=0]%
{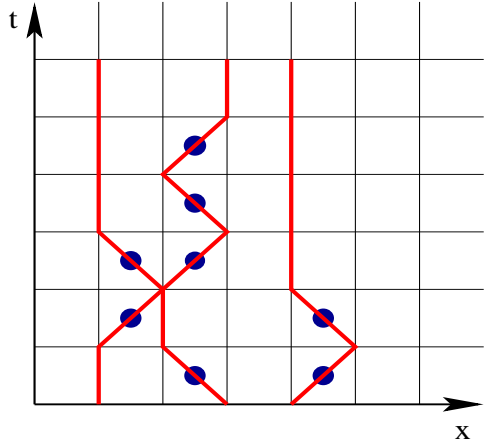}%
\caption{Shown are sample worldlines on a 1+1 dimensional lattice. The circles denote
a hop during a time step to a neighboring lattice site.}%
\label{worldlinefig}%
\end{center}
\end{figure}

We now wish to calculate the path integral by considering explicit fermion worldlines,
see Fig.~\ref{worldlinefig} for sample worldlines.

Clearly it is not feasible to take into account all possible worldlines even for 
smaller lattices, so we approximate the actual value of the path integral by
performing Monte Carlo sampling of the worldlines.
The contribution of the worldines to the path integral is most conveniently 
calculated in the occupation number basis which we will introduce in the following.

To this end, let us first simplify the situation by considering the action without 
spatial hopping and fixed
lattice site $\vec{n}_{s}$,
\begin{equation}
S^{free}_{\vec{n}_{s}}=c^{\ast}(n)c(n)-(1-6h)c^{\ast}(n)c(n-1) \ ,
\end{equation}
where $n$ is the temporal component of $\vec{n}$.
So we are considering the physics of a single free particle at zero temperature
with only one spatial lattice site, i.e., a $0+1$
dimensional system.
Next, let us define coherent quantum states
\begin{equation}
\left\vert \eta\right\rangle =e^{\eta a^{\dag}}\left\vert 0\right\rangle
=\left(  1+\eta a^{\dag}\right)  \left\vert 0\right\rangle ,
\end{equation}
where $\eta$ is a Grassmann variable.  Then%
\begin{equation}
a\left\vert \eta\right\rangle =ae^{\eta a^{\dag}}\left\vert
0\right\rangle = a\left(  1+\eta a^{\dag}\right)  \left\vert
0\right\rangle =\eta \left\vert
0\right\rangle 
\end{equation}
and%
\begin{align}
\left\langle \eta^{\prime}\right\vert \left.  \eta\right\rangle  &
=\left\langle 0\right\vert e^{\bar{\eta}^{\prime}a}e^{\eta a^{\dag}}\left\vert
0\right\rangle =\left\langle 0\right\vert (1+\bar{\eta}^{\prime}a)(1+\eta
a^{\dag})\left\vert 0\right\rangle \nonumber\\
&  =1+\bar{\eta}^{\prime}\eta=e^{\bar{\eta}^{\prime}\eta}.
\end{align}
Also, in this basis the identity is given by%
\begin{align}
\int d\bar{\eta}d\eta\,e^{-\bar{\eta}\eta}\left\vert \eta\right\rangle
\left\langle \eta\right\vert  &  =\int d\bar{\eta}d\eta\,\left(  1-\bar{\eta
}\eta\right)  (1+a^{\dag}\eta)\left\vert 0\right\rangle \left\langle
0\right\vert (1+\bar{\eta}a)\nonumber\\
&  =\int d\bar{\eta}d\eta(-\bar{\eta}\eta)\left\vert 0\right\rangle
\left\langle 0\right\vert +\int d\bar{\eta}d\eta\left(  \eta\bar{\eta}\right)
\left\vert 1\right\rangle \left\langle 1\right\vert \nonumber\\
&  =\left\vert 0\right\rangle \left\langle 0\right\vert +\left\vert
1\right\rangle \left\langle 1\right\vert =1 \ ,
\end{align}
where $\left\vert 1\right\rangle $ is the normalized state with occupation
number $1$.
We now do some reverse engineering. 
Let $O$ be the transition operator for the fermion from time slice $i$ to the next time slice
$i+1$.
We want the path integral with antiperiodic time boundary%
\begin{equation}
\int Dc \, Dc^* \exp\left(  -S^{free}_{\vec{n}_{s}} \right)
\end{equation}
to arise from the trace %or graded trace (for antiperiodic time boundary)
of an $n$-fold product $O^n$ where $n$ is the number of lattice sites in temporal
direction
\begin{align}  \label{eq:op}
&  \tr(O^N) = \int d\eta_{N-1}d\bar{\eta}_{N-1}e^{\bar{\eta}_{N-1}\eta_{N-1}}
\left\langle \eta_{N-1}\right\vert  O\cdot O\cdot O\cdots O   \left\vert \eta_{N-1}\right\rangle \nonumber\\
&  =\int d\eta_{N-1}d\bar{\eta}_{N-1}e^{\bar{\eta}_{N-1}\eta_{N-1}}
\left\langle \eta_{N-1}\right\vert O\int d\bar{\eta}%
_{N-2}d\eta_{N-2}e^{-\bar{\eta}_{N-2}\eta_{N-2}}\left\vert \eta_{N-2}\right\rangle
\nonumber\\[0.2cm]
& \times \left\langle \eta_{N-2}\right\vert O \int d\bar{\eta}%
_{N-3}d\eta_{N-3}e^{-\bar{\eta}_{N-3}\eta_{N-3}}\left\vert \eta_{N-3}\right\rangle
\left\langle \eta_{N-3}\right\vert O\ldots \quad \ldots \left\vert \eta_{0}\right\rangle
\left\langle \eta_{0}\right\vert O \left\vert \eta_{N-1}\right\rangle \ .
\end{align}
Explicit knowledge of the quantum operator $O$ is not necessary, but
its matrix elements are required to fulfill
\begin{equation}
\left\langle \eta_{i}\right\vert O\left\vert \eta_{i-1}\right\rangle
=\exp\left[  \bar{\eta}_{i}\eta_{i-1}(1-6h)\right] 
\end{equation}
such that the trace of $O^n$, Eq.~(\ref{eq:op}), reproduces the path integral.

Let us now calculate matrix elements of $O$ in the occupation number basis.
For brevity we merely present the calculation for 
$\left\langle 0\right\vert O\left\vert 0\right\rangle$, while the results for
the remaining matrix elements are deferred to the appendix, %
\begin{align}
\left\langle 0\right\vert O\left\vert 0\right\rangle  &  =\left\langle
0\right\vert \int d\bar{\eta}^{\prime}d\eta^{\prime}e^{-\bar{\eta}^{\prime
}\eta^{\prime}}\left\vert \eta^{\prime}\right\rangle \left\langle \eta
^{\prime}\right\vert O\int d\bar{\eta}d\eta e^{-\bar{\eta}\eta}\left\vert
\eta\right\rangle \left\langle \eta\right\vert \left.  0\right\rangle
\nonumber\\
&  =\int d\bar{\eta}^{\prime}d\eta^{\prime}\int d\bar{\eta}d\eta e^{-\bar
{\eta}\eta}e^{-\bar{\eta}^{\prime}\eta^{\prime}}\exp\left[  \bar{\eta}%
^{\prime}\eta (1-6h)\right]  \left\langle
0\right\vert \left.  \eta^{\prime}\right\rangle \left\langle \eta\right\vert
\left.  0\right\rangle \nonumber\\
&  =\int d\bar{\eta}^{\prime}d\eta^{\prime}\int d\bar{\eta}d\eta e^{-\bar
{\eta}\eta}e^{-\bar{\eta}^{\prime}\eta^{\prime}}\exp\left[  \bar{\eta}%
^{\prime}\eta (1-6h)\right] \nonumber\\
&  =\int d\bar{\eta}^{\prime}d\eta^{\prime}\int d\bar{\eta}d\eta(1-\bar{\eta
}\eta)(1-\bar{\eta}^{\prime}\eta^{\prime})\left[  1+\bar{\eta}^{\prime}\eta
(1-6h)\right] \nonumber\\
&  =\int d\bar{\eta}^{\prime}d\eta^{\prime}\int d\bar{\eta}d\eta(-\bar{\eta
}\eta)(-\bar{\eta}^{\prime}\eta^{\prime})=1.
\end{align}%

Next, we consider the case with more than one lattice site and with nearest-neighbor
hopping,%
\begin{align}
S^{free} &  =\sum_{\vec{n}}\left[  c^{\ast}(\vec{n})c(\vec{n})-
(1-6h)c^{\ast}(\vec{n})c(\vec{n}-\hat{0})\right]  \nonumber\\
&  -h \sum_{\vec{n},l_{s}}\left[  c^{\ast}%
(\vec{n})c(\vec{n}+\hat{l}_{s}-\hat{0})+c^{\ast}(\vec{n})c(\vec{n}-\hat
{l}_{s}-\hat{0})\right]  .
\end{align}
Let $\eta_{n-1}$, $\eta_{n}$ be the Grassmann variables at given spatial
lattice site (call it $A$) at time steps $n-1$ and $n.$  Let $\theta_{n-1}$,
$\theta_{n}$ be the Grassmann variables at a neighboring spatial lattice site
(call it $B$) at time steps $n-1$ and $n$. In order to reproduce the
path integral we demand%
\beqa
  \left\langle \eta_{n},\theta_{n}\right\vert O\left\vert \eta_{n-1}%
,\theta_{n-1}\right\rangle 
  &=& \exp\left[  \bar{\eta}_{n}\eta_{n-1}(1-6h)\right]  
 \exp\left[  \bar{\theta}_{n}\theta_{n-1}(1-6h)\right]  \nonumber \\
& \times&  \exp\left[  \bar{\eta}_{n}\theta_{n-1}h\right]  
 \exp\left[  \bar{\theta}_{n}\eta_{n-1}h\right]  .
\eeqa
We now construct the matrix elements of $O$ in the occupation number basis for
sites $A$ and $B$. To this aim, we consider the subspace when there is exactly one
particle in either $A$ or $B$. \ Let $\left\vert 1,0\right\rangle $ be the
state when the particle is at $A$. \ Let $\left\vert 0,1\right\rangle $ be the
state when the particle is at $B$. \ We find%

\beqa
 \left\langle 1,0\right\vert O\left\vert 1,0\right\rangle &  =&
 \int d\bar{\eta}^{\prime}d\eta^{\prime}\int d\bar{\eta}d\eta\int
d\bar{\theta}^{\prime}d\theta^{\prime}\int d\bar{\theta}d\theta
(1-\bar{\eta}\eta)(1-\bar{\eta}^{\prime}\eta^{\prime})(1-\bar{\theta
}\theta)(1-\bar{\theta}^{\prime}\theta^{\prime}) \nonumber\\
&  \times& \left[  1+\bar{\eta}^{\prime}\eta (1-6h)\right]  
\left[  1+\bar{\theta}^{\prime}\theta (1-6h)\right] \left[  1+\bar{\eta}^{\prime}\theta h\right]  
\left[  1+\bar{\theta}^{\prime}\eta h\right]  \eta^{\prime}\bar{\eta}\nonumber\\[0.2cm]
&  =& (1-6h).
\eeqa
Also%
\beqa
\left\langle 1,0\right\vert O\left\vert 0,1\right\rangle &  =&
\int d\bar{\eta}^{\prime}d\eta^{\prime}\int d\bar{\eta}d\eta\int
d\bar{\theta}^{\prime}d\theta^{\prime}\int d\bar{\theta}d\theta
(1-\bar{\eta}\eta)(1-\bar{\eta}^{\prime}\eta^{\prime})(1-\bar{\theta
}\theta)(1-\bar{\theta}^{\prime}\theta^{\prime})\nonumber\\
&  \times& \left[  1+\bar{\eta}^{\prime}\eta (1-6h)\right]  
\left[  1+\bar{\theta}^{\prime}\theta (1-6h)\right] \left[  1+\bar{\eta}^{\prime}\theta h\right]  
\left[  1+\bar{\theta}^{\prime}\eta h\right]  \eta^{\prime}\bar{\theta}\nonumber\\[0.2cm]
&  =&h.
\eeqa
Clearly by symmetry between sites $A$ and $B$ one has%
\begin{align}
\left\langle 0,1\right\vert O\left\vert 0,1\right\rangle  &  =(1-6h)\\
\left\langle 0,1\right\vert O\left\vert 1,0\right\rangle  &  =h.
\end{align}
The rules are now clear for a single free fermion worldline: for a hop to a
neighboring lattice site we get an amplitude $h$, 
whereas one obtains $(1-6h)$
if the particle stays at the same lattice site.

%%%%%%%%%%%%%%%%%%%%%%%%%%%%%%%%%%%%%%%%%%%%%%%%%%%%%%%%%%%%%%%%%%%%%%%%%%%%%%%%%
\subsection{Two-body interaction}

We continue by introducing a two-body interaction, the same for every pair of different
fermion species as required in the Wigner $SU(4$) limit. 
We take the two-body contact term of the Hamiltonian in Eq.~(\ref{eq:hamil})
\beq
H_2 = C_{0}\sum_{\vec{n}_{s},i>j}a_{i}^{\dagger}(\vec{n}_{s})a_{i}(\vec{n}%
_{s})a_{j}^{\dagger}(\vec{n}_{s})a_{j}(\vec{n}_{s}) \ .
\eeq
Utilizing a Hubbard-Stratonovich transformation with an auxiliary
scalar field $s(\vec{n})$ eliminates the four-fermion term and yields
\begin{align}
S &  =\sum_{\vec{n},i}\left[  c_{i}^{\ast}(\vec{n})c_{i}(\vec{n}+\hat
{0})-e^{\sqrt{-C_{0}\alpha_{t}}s(\vec{n})+\frac{C_{0}\alpha_{t}}{2}%
}(1-6h)c_{i}^{\ast}(\vec{n})c_{i}^{\prime}(\vec{n})\right]  \nonumber\\
&  -h\sum_{\vec{n},l_{s},i}\left[  c_{i}^{\ast}(\vec{n})c_{i}^{\prime}(\vec
{n}+\hat{l}_{s})+c_{i}^{\ast}(\vec{n})c_{i}^{\prime}(\vec{n}-\hat{l}%
_{s})\right]  +\frac{1}{2}\sum_{\vec{n}}s^{2}(\vec{n}).
\end{align}
The strength of the two-body coefficient $C_{0}$ is determined by summing
nucleon-nucleon bubble diagrams \cite{Lee:2004qd}.  Again we translate the
$c$ field%
\begin{align}
S &  =\sum_{\vec{n},i}\left[  c_{i}^{\ast}(\vec{n})c_{i}(\vec
{n})-e^{\sqrt{-C_{0}\alpha_{t}}s(\vec{n})+\frac{C_{0}\alpha_{t}}{2}%
}(1-6h)c_{i}^{\ast}(\vec{n})c_{i}(\vec{n}-\hat{0})\right]  \nonumber\\
&  -h\sum_{\vec{n},l_{s},i}\left[  c_{i}^{\ast}(\vec{n})c_{i}(\vec{n}+\hat
{l}_{s}-\hat{0})+c_{i}^{\ast}(\vec{n})c_{i}(\vec{n}-\hat{l}_{s}-\hat
{0})\right]  +\frac{1}{2}\sum_{\vec{n}}s^{2}(\vec{n})
\end{align}
and factorize the two-body interaction contribution%
\begin{align}  \label{eq:twobody}
\exp\left[  -S\right]   &  =\exp\left[  -S^{free}\right]  \nonumber\\
&  \times\prod\limits_{\vec{n}}\left[  e^{-\frac{1}{2}s^{2}(\vec{n})}%
\exp\left[  \left(  e^{\sqrt{-C_{0}\alpha_{t}}s(\vec{n})+\frac{C_{0}\alpha
_{t}}{2}}-1\right)  (1-6h)\sum_{i}c_{i}^{\ast}(\vec{n})c_{i}(\vec{n}-\hat
{0})\right]  \right]  .
\end{align}
For simplicity we start with the case when there are only two fermion species
which we call $i=\uparrow,\downarrow$. Expanding the exponential of the two-body interaction
in Eq.~(\ref{eq:twobody}) and making use of
\begin{equation}
\int ds(\vec{n})\;e^{-\frac{1}{2}s^{2}(\vec{n})}\left(  e^{\sqrt{-C_{0}%
\alpha_{t}}s(\vec{n})+\frac{C_{0}\alpha_{t}}{2}}-1\right)  =0
\end{equation}
we arrive at%
\begin{align}
&  \int ds(\vec{n})\;e^{-\frac{1}{2}s^{2}(\vec{n})}\exp\left[  \left(
e^{\sqrt{-C_{0}\alpha_{t}}s(\vec{n})+\frac{C_{0}\alpha_{t}}{2}}-1\right)
(1-6h)\sum_{i}c_{i}^{\ast}(\vec{n})c_{i}(\vec{n}-\hat{0})\right]  \nonumber\\
&  =\int ds(\vec{n})\left[  e^{-\frac{1}{2}s^{2}(\vec{n})}+e^{-\frac{1}%
{2}s^{2}(\vec{n})}\left(  e^{\sqrt{-C_{0}\alpha_{t}}s(\vec{n})+\frac
{C_{0}\alpha_{t}}{2}}-1\right)  ^{2}(1-6h)^{2}c_{\uparrow}^{\ast}(\vec
{n})c_{\uparrow}(\vec{n}-\hat{0})c_{\downarrow}^{\ast}(\vec{n})c_{\downarrow
}(\vec{n}-\hat{0})\right]  \nonumber\\[0.2cm]
%&  =\sqrt{2\pi}+\left(  e^{-C_{0}\alpha_{t}}\sqrt{2\pi}-2\sqrt{2\pi}%
%+\sqrt{2\pi}\right)  (1-6h)^{2}c_{\uparrow}^{\ast}(\vec{n})c_{\uparrow}%
%(\vec{n}-\hat{0})c_{\downarrow}^{\ast}(\vec{n})c_{\downarrow}(\vec{n}-\hat
%{0})\nonumber\\
%&  =\sqrt{2\pi}\left[  1+\left(  e^{-C_{0}\alpha_{t}}-1\right)  (1-6h)^{2}%
%c_{\uparrow}^{\ast}(\vec{n})c_{\uparrow}(\vec{n}-\hat{0})c_{\downarrow}^{\ast
%}(\vec{n})c_{\downarrow}(\vec{n}-\hat{0})\right]  \nonumber\\
&  =\sqrt{2\pi}\exp\left[  \left(  e^{-C_{0}\alpha_{t}}-1\right)
(1-6h)^{2}c_{\uparrow}^{\ast}(\vec{n})c_{\uparrow}(\vec{n}-\hat{0}%
)c_{\downarrow}^{\ast}(\vec{n})c_{\downarrow}(\vec{n}-\hat{0})\right]  .
\end{align}
The $\sqrt{2\pi}$ is an irrelevant factor which drops out of every physical
observable and will thus be neglected. The important term is%
\begin{equation}
\exp\left[  \left(  e^{-C_{0}\alpha_{t}}-1\right)  (1-6h)^{2}c_{\uparrow
}^{\ast}(\vec{n})c_{\uparrow}(\vec{n}-\hat{0})c_{\downarrow}^{\ast}(\vec
{n})c_{\downarrow}(\vec{n}-\hat{0})\right]  
\end{equation}
which yields a contribution only if an $\uparrow$ particle and a $\downarrow$
particle are at the same spatial site at the beginning of a time step and
neither one hops during the time step. For such a
configuration  the free particle action delivers an amplitude of $(1-6h)^{2}$
such that the entire amplitude including the interaction becomes%
\begin{equation}
(1-6h)^{2}+\left(  e^{-C_{0}\alpha_{t}}-1\right)  (1-6h)^{2}=e^{-C_{0}%
\alpha_{t}}(1-6h)^{2}\label{two-body} \ .%
\end{equation}
Hence there is an additional factor of $e^{-C_{0}\alpha_{t}}$ due to the two-body
interaction. The reason for this simple result is that when $h=0$, there is no
time discretization error in our lattice formulation.  It exactly matches the
Hamiltonian formulation without hopping.

We generalize our findings to the case with more than two fermion species. Expansion
of the two-body interaction term yields
\begin{align}
&  \exp\left[  \left(  e^{\sqrt{-C_{0}\alpha_{t}}s(\vec{n})+\frac{C_{0}%
\alpha_{t}}{2}}-1\right)  (1-6h)\sum_{i}c_{i}^{\ast}(\vec{n})c_{i}(\vec
{n}-\hat{0})\right]  \nonumber\\
&  =1+\left(  e^{\sqrt{-C_{0}\alpha_{t}}s(\vec{n})+\frac{C_{0}\alpha_{t}}{2}%
}-1\right)  (1-6h)\sum_{i}c_{i}^{\ast}(\vec{n})c_{i}(\vec{n}-\hat
{0})\nonumber\\
&  +\left(  e^{\sqrt{-C_{0}\alpha_{t}}s(\vec{n})+\frac{C_{0}\alpha_{t}}{2}%
}-1\right)  ^{2}(1-6h)^{2}\sum_{i>j}c_{i}^{\ast}(\vec{n})c_{i}(\vec{n}-\hat
{0})c_{j}^{\ast}(\vec{n})c_{j}(\vec{n}-\hat{0})\nonumber\\
&  +\left(  e^{\sqrt{-C_{0}\alpha_{t}}s(\vec{n})+\frac{C_{0}\alpha_{t}}{2}%
}-1\right)  ^{3}(1-6h)^{3}\sum_{i>j>k}c_{i}^{\ast}(\vec{n})c_{i}(\vec{n}%
-\hat{0})c_{j}^{\ast}(\vec{n})c_{j}(\vec{n}-\hat{0})c_{k}^{\ast}(\vec{n}%
)c_{k}(\vec{n}-\hat{0})\nonumber\\
&  +\cdots \ ,
\end{align}
where the ellipsis denotes higher orders in the Grassmann variables $c^*, c$.
The integral formula%
\begin{align}
\int ds(\vec{n})\;e^{-\frac{1}{2}s^{2}(\vec{n})}e^{k\left[  \sqrt{-C_{0}%
\alpha_{t}}s(\vec{n})+\frac{C_{0}\alpha_{t}}{2}\right]  } &  =\int ds(\vec
{n})\;e^{-\frac{1}{2}s^{2}(\vec{n})+k\sqrt{-C_{0}\alpha_{t}}s(\vec{n}%
)+k\frac{C_{0}\alpha_{t}}{2}}{}^{{}}\nonumber\\
&  =e^{(k-k^{2})\frac{C_{0}\alpha_{t}}{2}}\sqrt{2\pi}.
\end{align}
allows us to perform the integration over the auxiliary field $s$.
For three different fermion species, which we denote by $i=1,2,3$, we obtain after integration over $s$%
\begin{align}  \label{eq:3ferm}
&  \exp\left[  \left(  e^{-C_{0}\alpha_{t}}-1\right)  (1-6h)^{2}c_{1}^{\ast
}(\vec{n})c_{1}(\vec{n}-\hat{0})c_{2}^{\ast}(\vec{n})c_{2}(\vec{n}-\hat
{0})\right]  \nonumber\\
&  \times\exp\left[  \left(  e^{-C_{0}\alpha_{t}}-1\right)  (1-6h)^{2}%
c_{2}^{\ast}(\vec{n})c_{2}(\vec{n}-\hat{0})c_{3}^{\ast}(\vec{n})c_{3}(\vec
{n}-\hat{0})\right]  \nonumber\\
&  \times\exp\left[  \left(  e^{-C_{0}\alpha_{t}}-1\right)  (1-6h)^{2}%
c_{3}^{\ast}(\vec{n})c_{3}(\vec{n}-\hat{0})c_{1}^{\ast}(\vec{n})c_{1}(\vec
{n}-\hat{0})\right]  \nonumber\\
&  \times\exp\left[  \left(  e^{-3C_{0}\alpha_{t}}-3e^{-C_{0}\alpha_{t}%
}+2\right)  (1-6h)^{3}c_{1}^{\ast}(\vec{n})c_{1}(\vec{n}-\hat{0})c_{2}^{\ast
}(\vec{n})c_{2}(\vec{n}-\hat{0})c_{3}^{\ast}(\vec{n})c_{3}(\vec{n}-\hat
{0})\right]  .
\end{align}
Note that the apparent three-body interaction from the Hubbard-Stratonovich
transformation is $O(\alpha_{t}^{2})$, whereas the other terms in Eq.~(\ref{eq:3ferm})
are of order $O(\alpha_{t})$. 
It is therefore a discretization
effect which disappears as the temporal lattice spacing goes to zero. 

It is straightforward to see that this term 
simply produces the correct combinatorial factors in the pairwise
two-body interactions, and in fact is not a three-body interaction.
If there are exactly two particles of different species at the same site
and neither one hops during the time step, we get a contribution analogous
to Eq.~(\ref{two-body}),%
\begin{equation}
(1-6h)^{3}+\left(  e^{-C_{0}\alpha_{t}}-1\right)  (1-6h)^{3}=e^{-C_{0}%
\alpha_{t}}(1-6h)^{3} \ ,
\end{equation}
i.e., we obtain an extra factor of $e^{-C_{0}\alpha_{t}}$ from the two-body interaction.
But for three different particles at the same site and
no hop during the time step the contribution to the amplitude is%
\begin{align}
&  (1-6h)^{3}+3\left(  e^{-C_{0}\alpha_{t}}-1\right)  (1-6h)^{3}+\left(
e^{-3C_{0}\alpha_{t}}-3e^{-C_{0}\alpha_{t}}+2\right)  (1-6h)^{3}\nonumber\\
&  =e^{-3C_{0}\alpha_{t}}(1-6h)^{3}.
\end{align}
So we get a factor of $e^{-3C_{0}\alpha_{t}}$ from the three pairwise
two-body interactions. Again, the reason this is so simple is that when
$h=0$, there is no time discretization in our lattice formulation. Moreover, from
these considerations it should
be clear that for an arbitrary number of fermion species we get a factor
$e^{-C_{0}\alpha_{t}}$ for each pairwise interaction.

\subsection{Three-body interaction}

In the present investigation we choose the lattice three-body interaction in such a way
that for exactly three particles of different kind at the same site and no
hop during the time step the contribution to the amplitude reads
$ e^{-D_{0}\alpha_{t}}(1-6h)^{3}.\label{three-body}$
We assume that the lattice action with Hubbard-Stratonovich fields has been
designed appropriately to coincide with this definition. Its explicit form
is not relevant here.

\subsection{Results}

\begin{figure}
[ptb]
\begin{center}
\includegraphics[
height=3.in,
width=3.8in%,
%angle=-90
]%
{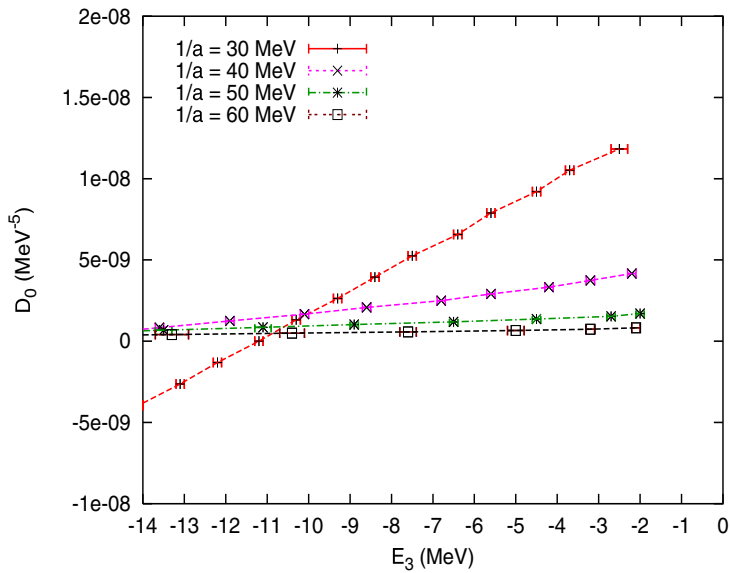}%
\caption{The three-body interaction $D_{0}$ as a function of the triton energy
$E_{3}$ for various lattice spacings using the Euclidean lattice formalism.
\ The two-nucleon binding energy is set at $1$~MeV.}%
\label{results}%
\end{center}
\end{figure}

After having discussed the possible contributions to the worldlines of three distinct fermions
on the lattice, we present now the results of the simulation.
Let $\left\vert 0,0,0\right\rangle $ be the state with three nucleons, each of different kind and 
each with zero momentum. We employ path integral Monte Carlo with worldlines to
compute lattice approximations for
\begin{equation}
G(\beta)=\left\langle 0,0,0\right\vert \exp\left[  -\beta H\right]  \left\vert
0,0,0\right\rangle \ ,
\end{equation}
making use of the fact that the state $\left\vert 0,0,0\right\rangle $ has a non-vanishing overlap with
the three-body ground state of the Hamiltonian $H$. 
By measuring the exponential tail of $G(\beta)$ for large $\beta$, we have a
measurement for the triton ground state energy $E_{3}$. \ For pairs of spatial
and temporal lattice spacings,%
\begin{align}
a^{-1}  & =30\text{\ MeV},\text{ }a_{t}^{-1}=9\text{\ MeV;}\\
a^{-1}  & =40\text{\ MeV},\text{ }a_{t}^{-1}=16\text{\ MeV;}\\
a^{-1}  & =50\text{\ MeV},\text{ }a_{t}^{-1}=25\text{\ MeV;}\\
a^{-1}  & =60\text{\ MeV},\text{ }a_{t}^{-1}=36\text{\ MeV;}%
\end{align}
we compute $D_{0}$ as a function of the triton energy $E_{3}$. \ The results
are shown in Fig.~\ref{results}.%

%TCIMACRO{\FRAME{ftbpFU}{2.981in}{4.2436in}{0pt}{\Qcb{The three-body
%interaction $D_{0}$ as a function of the triton energy $E_{3}$ for various
%lattice spacings using the Euclidean lattice formalism. \ The two-nucleon
%binding energy is set at $1$ MeV.}}{\Qlb{results}}{results.eps}%
%{\special{ language "Scientific Word";  type "GRAPHIC";
%maintain-aspect-ratio TRUE;  display "USEDEF";  valid_file "F";
%width 2.981in;  height 4.2436in;  depth 0pt;  original-width 4.8948in;
%original-height 6.9877in;  cropleft "0";  croptop "1";  cropright "1";
%cropbottom "0";  filename 'results.eps';file-properties "XNPEU";}}}%
%BeginExpansion
%EndExpansion
We have stayed away from the sensitive region $E_{3}\approx-1\ $MeV. \ Since
there are many dimer plus nucleon continuum states near the energy threshold
$-1$ MeV, it is difficult to extract $E_{3}$ accurately. It requires lattices with larger
temporal extensions
and significantly more CPU time to study this critical region. For $E_{3}<-2$
MeV, however, the simulation is not difficult and the finite volume effects
are small. \ For $a^{-1}=30$\ MeV and $40$\ MeV we used a lattice volume of
$L^3=6^3$; for $a^{-1}=50$\ MeV we used $L^3=7^3;$ and for $a^{-1}=60$\ MeV we used
$L^3=8^3$. \ The error bars shown are stochastic error estimates, determined by
the data from four different processors with completely independent runs.
\ The finite volume errors are significantly smaller than the stochastic error
estimates shown. \ We note the clear similarity between the Euclidean lattice
results and Hamiltonian lattice results.

%%%%%%%%%%%%%%%%%%%%%%%%%%%%%%%%%%%%%%%%%%%%%%%%%%%%%%%%%%%%%%%%%%%%%%%%%%%%%%%%
\section{Lattice action positivity}

\label{sec:hubstr}

Pure neutron matter has been studied on the lattice using effective field
theory both with pions \cite{Lee:2004si} and without pions
\cite{Lee:2004qd,Lee:2005is, Lee:2005it}. \ For cold dilute neutron matter at
momentum scales below the pion mass, pionless effective field theory should
provide an adequate description of the low-energy physics. \ One nice feature
of the lowest-order pionless theory for pure neutron matter is that it can be
implemented on the lattice with a positive semi-definite action by means of a
Hubbard-Stratonovich transformation. \ This feature is very important for
performing efficient Monte Carlo simulations.

For cold dilute nuclear matter with a small proton fraction, one expects that
pionless effective field theory also describes the relevant low-energy
physics. \ However, in this case a three-nucleon force is required for
consistent renormalization. \ Until recently it has been an open question
whether or not this three-nucleon force would spoil positivity of the lattice
action. \ It was shown in \cite{Chen:2004rq} that in the Wigner
SU(4)-symmetric limit, the lattice action would remain positive so long as the
three-nucleon force was not too strong and the four-nucleon force was not too
repulsive. \ However, it was not known if the parameters of the real world
satisfy these conditions. \ In this section we show that these conditions
are indeed satisfied for typical lattice spacings relevant for cold dilute
nuclear matter.

Consider a Euclidean lattice action with two-, three- and four-body
interactions of the form
\begin{equation}
\sum_{\vec{n}}\left[  c_{2}(\bar{\psi}\psi(\vec{n}))^{2}+c_{3}(\bar{\psi}%
\psi(\vec{n}))^{3}+c_{4}(\bar{\psi}\psi(\vec{n}))^{4}\right]  \ .
\end{equation}
In \cite{Chen:2004rq} it was shown that a Wigner SU(4)-symmetric nuclear
lattice simulation is possible with unequal numbers of protons and neutrons
without a sign problem if and only if the matrix%
\begin{equation}%
\begin{bmatrix}
1 & 0 & 2c_{2}\\
0 & 2c_{2} & 6c_{3}\\
2c_{2} & 6c_{3} & 12c_{2}^{2}+24c_{4}%
\end{bmatrix}
\end{equation}
is positive semi-definite, with the added condition that if $c_{2}=0$ then
$c_{4}=0$. \ The determinant of this matrix is $16c_{2}^{3}-36c_{3}%
^{2}+48c_{2}c_{4}$. \ The relations between $c_{2}$ and $c_{3}$ (in lattice
units) and our parameters $C_{0}$ and $D_{0}$ (in physical units) are%
\begin{align}
c_{2}  &  =\frac{1}{2}(1-6h)^{2}\left(  e^{-C_{0}a^{-3}a_{t}}-1\right)  ,\\
c_{3}  &  =\frac{1}{6}(1-6h)^{3}\left(  e^{-D_{0}a^{-6}a_{t}}-1\right)  .
\end{align}
While the three-nucleon force is required for consistent renormalization at
lowest order, the four-nucleon force is an irrelevant operator and can be
neglected for sufficiently small lattice spacings \cite{Platter:2004zs}. \ If we assume the
four-nucleon force to be zero, then the determinant is $16c_{2}^{3}%
-36c_{3}^{2}$.

For an SU(4)-symmetric triton binding energy of $8$ MeV, we find the results
shown in Table~\ref{table}.%
\begin{table}
\centering
\begin{tabular}
[c]{|c|c|c|c|c|c|c|c|}\hline
$a^{-1}$ (MeV)& $a_{t}^{-1}$ (MeV)& $h$ & $C_{0}$ (MeV$^{-2}$) & $D_{0}$ (MeV$^{-5}$) &
$c_{2}$ & $c_{3}$ & $16c_{2}^{3}-36c_{3}^{2}$\\\hline
$30$ & $9$ & $0.0532$ & $-2.66\times10^{-4}$ & $4.52\times10^{-9}$ & $0.28$ &
$-0.016$ & $0.35$\\\hline
$40$ & $16$ & $0.0532$ & $-1.83\times10^{-4}$ & $2.22\times10^{-9}$ & $0.25$ &
$-0.023$ & $0.23$\\\hline
$50$ & $25$ & $0.0532$ & $-1.39\times10^{-4}$ & $1.09\times10^{-9}$ & $0.23$ &
$-0.026$ & $0.18$\\\hline
$60$ & $36$ & $0.0532$ & $-1.13\times10^{-4}$ & $5.64\times10^{-10}$ & $0.22$
& $-0.027$ & $0.15$\\\hline
\end{tabular}
\caption{Shown are the values for $c_{2}, c_{3}$ and the determinant $16c_{2}^{3}-36c_{3}^{2}$
           for given values of the lattice spacings $a$ and $a_t$.}
\label{table}
\end{table}
We see that for lattice spacings in this range the determinant $16c_{2}%
^{3}-36c_{3}^{2}$ remains positive. \ This implies that many-body simulations
of cold dilute nuclear matter in the Wigner symmetry limit should be possible
without a sign problem, even for unequal numbers of protons and neutrons.

As noted above, we have ignored the irrelevant operator producing an
SU(4)-symmetric four-nucleon force. \ If we do include a small four-nucleon
force, this positivity result should most likely remain intact. \ However, this
will be addressed explicitly in a future study of the four-nucleon system
\cite{BKLM}.

%%%%%%%%%%%%%%%%%%%%%%%%%%%%%%%%%%%%%%%%%%%%%%%%%%%%%%%%%%%%%%%%%%%%%%%%%%%%%%%%
\section{Conclusions}  \label{sec:concl}

In this work we have introduced two novel approaches to few-body nuclear lattice simulations.
We worked in the Wigner SU(4)-symmetric limit where the $S$-wave scattering
lengths are equal.
First, in the Hamiltonian lattice formalism with the Lanczos method  
the three-body interaction strength was calculated as a function of 
the triton binding energy for various lattice spacings, while keeping the two-nucleon
binding energy fixed at $1$ MeV. Our findings indicate
that the coupling scales roughly---as expected---as the lattice spacing squared. 
We furthermore observe a pole singularity in the coupling slightly above
the threshold value of $1$ MeV in the triton binding energy. 
This is in good agreement with continuum regularization
results.

The second framework applied is  the Euclidean lattice formalism with
Monte Carlo sampling of the path integral. We observe a clear similarity
with the Hamiltonian lattice results for triton binding energies greater than
$2$ MeV. For binding energies closer to the threshold energy of $1$~MeV
it is difficult to extract the binding energy accurately with this method
due to the presence of many dimer plus nucleon continuum states in this energy region.

Although we have restricted ourselves to three nucleons and two- and three-nucleon forces
in the present investigation,
our results suggest that the Euclidean lattice method can be generalized to a larger
number of nucleons and more complicated forces amongst them. 
It sets the stage for further studies in the few-body sector,
such as the inclusion of a four-body force which can be easily included in this 
formalism or effects due to breaking of Wigner symmetry \cite{BKLM}.
The importance of higher chiral orders in the action can be studied as well.

Moreover, this work provides a first measurement of the three-nucleon force
on the lattice in the Wigner symmetry limit and is an important step
towards future many-body simulations with arbitrary numbers of nucleons including
three-body effects.

%%%%%%%%%%%%%%%%%%%%%%%%%%%%%%%%%%%%%%%%%%%%%%%%%%%%%%%%%%%%%%%%%%%%%%%%%%%%%%%%
\section*{Acknowledgments}
We thank H.-W. Hammer for useful discussions and reading the manuscript.
Partial financial support of Deutsche Forschungsgemeinschaft, Forschungszentrum
J\"ulich and U.S. Department of Energy (grant DE-FG02-04ER41335) is gratefully acknowledged.
This research is part of the EU Integrated Infrastructure Initiative Hadronphysics under contract
 number RII3-CT-2004-506078. D.~L. also thanks for the 
kind hospitality during his stay in J\"ulich where part of this work was carried out.

%%%%%%%%%%%%%%%%%%%%%%%%%%%%%%%%%%%%%%%%%%%%%%%%%%%%%%%%%%%%%%%%%%%%%%%%%%%%%%%%
\appendix

%%%%%%%%%%%%%%%%%%%%%%%%%%%%%%%%%%%%%%%%%%%%%%%%%%%%%%%%%%%%%%%%%%%%%%%%%%%%%%%%
\section{Matrix elements of $\mbox{\boldmath$O$}$}  \label{app}
%%%%%%%%%%%%%%%%%%%%%%%%%%%%%%%%%%%%%%%%%%%%%%%%%%%%%%%%%%%%%%%%%%%%%%%%%%%%%%%%

In this appendix, we present the remaining matrix elements of $O$ in the occupation 
number basis not given in the main text. For a lattice with one spatial lattice site
and one fermion we obtain
\begin{align}
\left\langle 0\right\vert O\left\vert 1\right\rangle  &  =\left\langle
0\right\vert \int d\bar{\eta}^{\prime}d\eta^{\prime}e^{-\bar{\eta}^{\prime
}\eta^{\prime}}\left\vert \eta^{\prime}\right\rangle \left\langle \eta
^{\prime}\right\vert O\int d\bar{\eta}d\eta e^{-\bar{\eta}\eta}\left\vert
\eta\right\rangle \left\langle \eta\right\vert \left.  1\right\rangle
\nonumber\\
&  =\int d\bar{\eta}^{\prime}d\eta^{\prime}\int d\bar{\eta}d\eta e^{-\bar
{\eta}\eta}e^{-\bar{\eta}^{\prime}\eta^{\prime}}\exp\left[  \bar{\eta}%
^{\prime}\eta (1-6h)\right]  \bar{\eta}\nonumber\\
&  =0
\end{align}%
\begin{align}
\left\langle 1\right\vert O\left\vert 0\right\rangle  &  =\left\langle
1\right\vert \int d\bar{\eta}^{\prime}d\eta^{\prime}e^{-\bar{\eta}^{\prime
}\eta^{\prime}}\left\vert \eta^{\prime}\right\rangle \left\langle \eta
^{\prime}\right\vert O\int d\bar{\eta}d\eta e^{-\bar{\eta}\eta}\left\vert
\eta\right\rangle \left\langle \eta\right\vert \left.  0\right\rangle
\nonumber\\
&  =\int d\bar{\eta}^{\prime}d\eta^{\prime}\int d\bar{\eta}d\eta e^{-\bar
{\eta}\eta}e^{-\bar{\eta}^{\prime}\eta^{\prime}}\exp\left[  \bar{\eta}%
^{\prime}\eta (1-6h)\right]  \eta^{\prime
}\nonumber\\
&  =0
\end{align}%
\begin{align}
\left\langle 1\right\vert O\left\vert 1\right\rangle  &  =\left\langle
1\right\vert \int d\bar{\eta}^{\prime}d\eta^{\prime}e^{-\bar{\eta}^{\prime
}\eta^{\prime}}\left\vert \eta^{\prime}\right\rangle \left\langle \eta
^{\prime}\right\vert O\int d\bar{\eta}d\eta e^{-\bar{\eta}\eta}\left\vert
\eta\right\rangle \left\langle \eta\right\vert \left.  1\right\rangle
\nonumber\\
&  =\int d\bar{\eta}^{\prime}d\eta^{\prime}\int d\bar{\eta}d\eta e^{-\bar
{\eta}\eta}e^{-\bar{\eta}^{\prime}\eta^{\prime}}\exp\left[  \bar{\eta}%
^{\prime}\eta (1-6h)\right]  \eta^{\prime}\bar{\eta
}\nonumber\\
&  =\int d\bar{\eta}^{\prime}d\eta^{\prime}\int d\bar{\eta}d\eta(1-\bar{\eta
}\eta)(1-\bar{\eta}^{\prime}\eta^{\prime})\left[  1+\bar{\eta}^{\prime}\eta
(1-6h)\right]  \eta^{\prime}\bar{\eta}\nonumber\\
&  =\int d\bar{\eta}^{\prime}d\eta^{\prime}\int d\bar{\eta}d\eta\left[
\bar{\eta}^{\prime}\eta\eta^{\prime}\bar{\eta}
(1-6h)\right]  =(1-6h).
\end{align}

\end{document}